\input epsf.sty
\newdimen\psfigsize
\def\psfigure#1 #2 #3 #4 #5{
    \begin{figure}[tbh]
      \vbox{
        \null\vskip-0.2in\hskip#2
        \epsfxsize=#1
        \epsfbox{#4}
        \vskip -0.3in
        \caption {#5 \label{#3}}
        \vskip 0.0truein plus0.2truein
      }
    \end{figure}
}
\newcommand{\beq}{\begin{equation}}
\newcommand{\eeq}{\end{equation}}
\newcommand{\bea}{\begin{eqnarray}}
\newcommand{\eea}{\end{eqnarray}}

\def\text#1{{\rm #1}}

\documentstyle[twoside,fleqn,espcrc2]{article}
\begin{document}
\title{Dynamical Fermion on Random-Block Lattice}
\author{Ting-Wai Chiu%
\address{Department of Physics, National Taiwan University,
Taipei, Taiwan 106, R.O.C.} \thanks{%
This work is supported by the National Science Council, R.O.C. under the
grant number NSC84-2112-M002-C13 and NSC85-2112-M002-013.}}

\begin{abstract}
Massless fermion field interacting with abelian dynamical gauge field on
2-dimensional random-block lattices are investigated using Hybrid Monte Carlo
simulations. Preliminary results of the Wilson loop and the chiral correlation
function are in agreement with the continuum Schwinger model.
\end{abstract}

\maketitle

\section{INTRODUCTION}

The problem of formulating the chiral fermion field on space-time lattices
has been discussed from the viewpoint of funtional integrals \cite{chiu95}.
I argued that the functional integral measure provided by any $single$ lattice
( hypercubical lattice or random lattice ) is not sufficent to yield fermionic
functional integrals to agree with the continuum field theory. My proposal of
using an ensemble of random-block lattices ( RBL ) to provide a proper measure
has been investigated. RBL regularization has been tested
for the free fermion ( massive and massless )
in $2d$ and $4d$ \cite{chiu88,chiu89} respectively, and for $2d$ massless
fermion in an external abelian gauge field \cite{chiu93,chiu94}. All
observables including fermion propagators, gauge invariant currents, axial
anomaly and current-current correlations are all in good agreement with the
continuum field theory. It is natural for us to proceed to investigations
involving dynamical fermion field on RBL. In this paper, massless fermion field
interacting with the dynamical abelian gauge field on 2d RBL is studied.
Using Hybrid Monte Carlo simulations, the Wilson loop and the chiral
correlation function are measured and preliminary results of these observables
are in agreement with the continuum Schwinger model. More extensive
measurements involving higher statistics and larger lattices are now in
progress and will be reported elsewhere.

\section{SCHWINGER MODEL on RBL}

The action of a massless fermion field interacting with the dynamical
abelian gauge field on a $2d$ RBL is
\begin{eqnarray*}
  A_{rbl} = \sum_{i,\mu} \omega_{i} K_{i}^{\mu }
            {\bar{\psi}_{i}} \gamma_{\mu}
            [ U_{\mu }(i) \psi_{i+\hat{\mu}}
            - U_{\mu }^{\dagger}(i-\hat{\mu}) \psi_{i-\hat{\mu}} ] \\
            + \beta \sum_{p} w_{p} [1-\text{Re}(U_{p})]
\label{eq:sch_rbl}
\end{eqnarray*}
where $\psi_i$ and $\bar{\psi}_i$ are two independent 2-component
spinors at site $i$,
$ \omega_i = \prod_\mu \frac{( x_{i+{\hat{\mu}}} - x_{i-{\hat{\mu}}})}{2} $
is the site weight at site $i$,
$ K_{i}^{\mu } = (x_{i+{\hat{\mu}}}-x_{i-{\hat{\mu}}})^{-1} $
is the inverse of the distance between the sites $i+\hat{\mu}$ and
$i-\hat{\mu}$ and
$ U_{\mu}(i)=\exp [\text{ie}\int_{x_i}^{x_{i+\mu}}\vec{A}\cdot\vec{dl}] $
is the link variable pointing from the site $i$ to the site $i+\hat{\mu}$,
and the $ \gamma $ matrices are chosen to be $\gamma_1=\sigma_1$,
$\gamma_2=\sigma_2$ and $\gamma_5=-i\gamma_1 \gamma_2 =\sigma_3 $.
The last term in the action $ A_{rbl} $ is due to the pure gauge field and
is conveniently labeled as
\begin{eqnarray*}
A_g[U] = \beta \sum_p w_{p} [1-\text{Re}(U_{p})]
\end{eqnarray*}
where $ \beta=\frac{1}{e^{2}} $, $ U_p $ is the path-ordered product of the
link variables around a plaquette $p$, $ w_p = \frac{1}{S_{p}} $ is the
weight of the plaquette $p$ and $ S_p $ the area of the plaquette.
The action $ A_{rbl} $ is invariant under the $U_{V}(1)$ transformation
\begin{eqnarray*}
\psi_i &\rightarrow& \exp ( \text{-ie} \theta_{i}) \psi_i \\
\bar{\psi}_i &\rightarrow& \bar{\psi}_i \exp ( \text{ ie} \theta_i) \\
U_{\mu}(i) &\rightarrow& \exp ( \text{-ie} \theta_i ) U_{\mu}(i)
                         \exp ( \text{ ie} \theta_{i+\hat{\mu}} )
\end{eqnarray*}
and the $U_{A}(1)$ transformation
\begin{eqnarray*}
\psi_i &\rightarrow& \exp ( \text{ ie}\gamma _{5} \alpha ) \psi_i \\
\bar{\psi}_i &\rightarrow& \bar{\psi}_i
                           \exp ( \text{ ie} \gamma_{5} \alpha )
\end{eqnarray*}
where $ \alpha $ is a global parameter. In the continuum limit,
$ A_{rbl} $ goes to the Schwinger model \cite{sch62}
\begin{eqnarray*}
A_{S}=\int d^2 x \left( {\bar{\psi}}(x) \gamma_\mu
        [ \partial_\mu + \text{ie} A_\mu (x)] \psi (x)
        + \frac{1}{4} F_{\mu\nu}^2 \right)
\end{eqnarray*}
The partition function of $A_{rbl}$ is
\begin{eqnarray*}
Z &=& \int[dU][d\psi][d\bar{\psi}] \exp ( - A_{rbl} ) \\
  &=& \int[dU] det(G) \exp(-A_g[U])
\end{eqnarray*}
where $G$ is the matrix
\begin{eqnarray*}
G_{ij}^{\alpha\beta} = \gamma_\mu^{\alpha \beta } \omega_i K_i^{\mu}
 [  U_\mu(i) \delta_{j,i+\hat{\mu}}
  - U_\mu^{\dagger}(i-\hat{\mu}) \delta_{j,i-\hat{\mu}} ]
\end{eqnarray*}
and $ det(G) $ is {\it positive definite even for one single flavor
of the fermion field }.
With $\gamma_1=\sigma_1$, $ \gamma_2=\sigma_2$, $ G $ can be written
as
\begin{eqnarray*}
G=\left( \begin{array}{cc} 0 & T_{1}-iT_{2} \\
                           T_{1}+iT_{2} & 0
         \end{array}
  \right)
\end{eqnarray*}
where $ T_1 $ and $ T_2 $ are anti-hermitian matrices,
$ T_{1}^{\dagger } = -T_{1} $ and $ T_{2}^{\dagger } = -T_{2} $. Then
\begin{eqnarray*}
G &=& \left( \begin{array}{cc}      0       & -(T_{1}+iT_{2})^{\dagger} \\
                                T_{1}+iT_{2} &        0
             \end{array}
      \right)
\end{eqnarray*}
and
\begin{eqnarray*}
\det G &=& \det (T_{1}+iT_{2}) \cdot
           \det (T_{1}+iT_{2})^{\dagger}(-1)^{n_{s}} \\
       &=& \left| \det (T_{1}+iT_{2}) \right|^{2} \\
     & \equiv & \det (K^{\dagger}K) > 0
\end{eqnarray*}
where $ K \equiv T_1 + i T_2 $ and $ n_{s} $ is the total number of sites
which is assumed to be even. Therefore the probability distribution
function is positive definite for any gauge configurations and thus amenable
for Monte Carlo simulations. Complex scalar fields
\{$\phi_i,\phi_i^{*}$\} are introduced to convert $ det(G) $ to a
functional integral over the complex scalar fields. The partition function
becomes
\begin{eqnarray*}
Z \simeq \int [dU] [d\phi^{\dagger}] [d\phi]
  \exp \left( -A_g[U] - \phi^{\dagger}(K^{\dagger}K)^{-1} \phi \right) \\
  \simeq  \int [dU] [d\xi^{\dagger}] [d\xi]
  \exp \left( -A_g[U] -\xi^{\dagger} \xi \right)
\end{eqnarray*}
where $ \xi  \equiv  (K^{\dagger })^{-1}\phi $. Then the system is equivalent
to that of effective action $ A_{eff} =  A_g[U] + \xi^{\dagger} \xi $ in
which quantum expectation value of any observables can be computed by the
Hybrid Monte Carlo ( HMC ) simulation \cite{hmc87}. The procedures of
performing the HMC simulation are similar to those on a hypercubical lattice.

\section{OBSERVABLES}

Measurements of the Wilson loop and the chiral correlation
function at $ \beta = 10.0 $ were performed using an ensemble of $64$
random-block lattices each of size $12 \times 12$ and average lattice
spacing $a=1.0$. On each RBL, $2000$ sweeps are used for thermalization,
$1000$ measurements were performed with consecutive measurements separated
by $10$ sweeps. More extensive measurements with higher statistics and on
larger lattices are now in progress and will be presented elsewhere.

\subsection*{CHIRAL CORRELATION FUNCTION}

One remarkable feature of the continuum Schwinger model is that all
gauge-invariant quantities can be computed exactly. The most interesting
physics of the model is that the chiral symmetry is dynamically broken
by quantum corrections. This can be observed in the chiral correlation function
\[
C(x,y) \equiv  < \bar{\psi}(x) \frac{1+\gamma_5}{2} \psi(x)
                 \bar{\psi}(y) \frac{1-\gamma_5}{2} \psi(y) >
\]
which goes to a non-vanishing constant
% $   frac{ {\text{e}}^2 } {  8 \pi^3 } \exp ( 2 \gamma_E ) $
for large $ | x - y | $, directly indicating the breakdown of clustering and
degenerate vacua in the theory. In a finite square of size $ L \times L $,
$ C(x,y) $ can be evaluated in the following.
\[
   C(x,y) = \left[ \sum_{\mu} S_{\mu}^{2}(x,y) \right] \times
\]
\[
   \exp [ 4 \pi ( \Delta_{0}(x,x)-\Delta(x,x)-\Delta_{0}(x,y)+\Delta(x,y) ) ]
\]
where $S_{\mu}(x,y)$ are components of the massless fermion propagator,
$\Delta_{0}(x,y)$ is the massless scalar propagator and
$\Delta(x,y)$ is the massive scalar propagator with mass
$\sqrt{\frac{{\rm e}^{2}}{\pi}}$. They can be expressed in the following.
\begin{eqnarray*}
S_{F}(x,y) = \gamma_\mu S_\mu(x,y)=\gamma_\mu \partial_\mu \Delta_{F}(x,y)
\end{eqnarray*}
where
\[
\Delta_{F}(x,y) = \frac{1}{L^2} \sum_{n_1,n_2=-\infty}^{\infty}
 \frac{ \exp [ i \pi ( \frac{ 2 \vec{n}+1}{L} ) \cdot ( \vec{x}-\vec{y} ) ] }
      { [ \frac{(2n_1+1) \pi }{L} ]^2 + [ \frac{(2n_2+1) \pi }{L} ]^2 }
\]
\[
\Delta_{0}(x,y) = \frac{1}{L^{2}} \sum_{n_1,n2=-\infty}^{\infty}
\frac{\exp [ i\ \frac{2\pi }{L} \vec{n} \cdot ( \vec{x}- \vec{y} ) ]}
     { ( \frac{2n_{1}\pi }{L} )^2 + ( \frac{2n_{2} \pi }{L} )^2 }
\]
\[
\Delta (x,y) = \frac{1}{L^{2}} \sum_{n_1,n_2=-\infty}^{\infty}
 \frac{\exp [ i \frac{2\pi}{L} \vec{n} \cdot ( \vec{x}-\vec{y}) ] }
 { ( \frac{2n_{1} \pi}{L} )^{2} + ( \frac{2n_{2} \pi}{L} )^{2} +
   \frac{{\rm e}^{2}}{\pi} }
\]
On a RBL, the chiral correlation function can be measured as
\[
   C_l(x,y)=\frac{1}{Z} \int [dU] \ det(G) \ \exp ( -A_g[U] ) \times
\]
\[
   \qquad \qquad \text{ (-1) Tr} [ G^{-1}(x,y) \frac{1-\gamma_{5}}{2}
                                    G^{-1}(y,x) \frac{1+\gamma_{5}}{2} ]
\]
\[
= \frac{1}{Z} \int [dU] [d\xi][d\xi^{\dagger}]
  | K^{-1}(x,y) |^{2} \exp ( -A_g[U] - \xi^{\dagger} \xi )
\]
where $ K^{-1}(x,y) $ is the right-handed fermion propagator in background
gauge field and complex scalar field, which can be computed in the following.
\[
  K_{i j}^{-1}
 = [(K^{\dagger })^{-1}K^{\dagger}K]_{i j}^{-1}
 = Q_{i l}^{-1} K_{ l j }^{\dagger}
\]
where $ Q \equiv K^{\dagger} K $ and the last step is computed by the
conjugate gradient method.
The ensemble average is obtained by repeating the same calculation for
each RBL while holding the end-points $ x $ and $ y $ fixed in all RBL,
\[
<C(x,y)> = \frac{1}{V^{N_{s}-2}} \int \prod_{i=1}^{N_{s}} d^{2}z_{i}
\]
\[
   \qquad \qquad \qquad \qquad \qquad   \delta (z_1-x) \delta (z_2-y) C(x,y)
\]
\[
 \simeq \frac{1}{N_{latt}} \sum_{l=1}^{N_{latt}} C_{l} (x,y)
\]
This RBL regularized chiral correlation function $ < C(x,y) > $ is
then compared to the continuum result. In Fig.~\ref{fig:cc},
$ <C(x,y)> $ is plotted for an ensemble of 64 RBL each of size
$12 \times 12$ and average lattice spacing $a=1.0$. It agrees with
the continuum Schwinger model and goes to a non-vanishing constant
at large $ | x - y | $.
\psfigure 3.0in -0.2in {fig:cc} {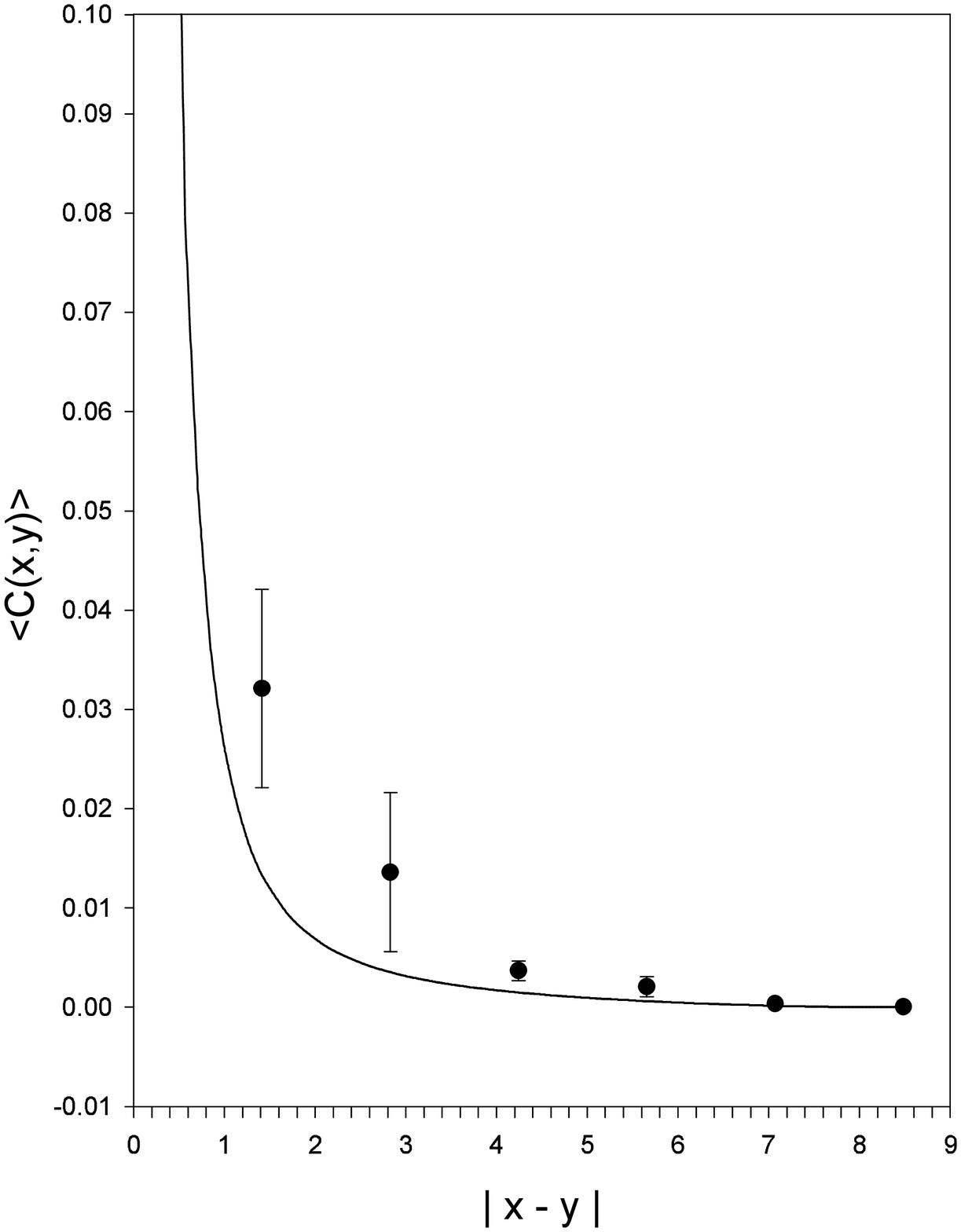} {
The chiral correlation function \ $ <~C~(x,y)~> $ plotted vs. $ |x-y| $
at $ \beta=10.0 $. The solid line corresponds to the continuum Schwinger
model.
}

\subsection*{WILSON LOOP}

The chiral symmetry breaking in the Schwinger model results in complete
screening of the linear confinement potential of external charges in the
pure gauge theory, and the quantum expectation of the Wilson loop
$ W(C) = \exp \left( i\text{e} \oint_{C} \vec{A} \cdot d \vec{x} \right) $
would display asymptotically perimeter law rather than the area law. In
continuum,
\begin{eqnarray*}
<W(C)> = \exp \left[ -\frac{\text{e}^{2}}{4\pi}
                 \oint_{C} \oint_{C} dx_\mu dy^\mu K_0 ( m |x-y| ) \right]
\end{eqnarray*}
where $ m = \sqrt{\frac{\text{e}^{2}}{\pi}} $.
On a RBL, the Wilson loop can be measured as
\begin{eqnarray*}
W(C)_l = \frac{\int [dU][d\psi][d\bar{\psi}] W(C) \exp ( -A_{rbl} ) }
                 {\int [dU][d\psi][d\bar{\psi}] \exp ( -A_{rbl} ) } \\
= \frac{\int[dU][d\xi][d\xi^{\dagger}] W(C) \exp(-A_g[U]-\xi^{\dagger}\xi)}
          {\int[dU][d\xi][d\xi^{\dagger}] \exp(-A_g[U]-\xi^{\dagger}\xi)}
\end{eqnarray*}
The ensemble average can be obtained by repeating the same calculation for
each RBL while holding the boundary of the Wilson loop fixed in all RBL.
For a square Wilson loop of size $ D \times D $, in practice, we can increase
the statistics by measuring all Wilson loops having $ D $ links in both
directions respectively, and the average area of these Wilson loops is
$D \times D$. Unlike the chiral correlation function or other
observables involving fermion fields, the fluctuations of the Wilson loop
from one RBL to another is rather small and the number of RBL used for
ensemble averaging can be minimum. In Fig.~\ref{fig:wloop}, the quantum
expectation values of $ D \times D $ Wilson loops for an ensemble of 64 RBL
are plotted vs. $ D $. The values are in good agreement with the
continuum Schwinger model, indicating the complete screening of the
confinment potential in the pure gauge theory.

\psfigure 3.0in -0.2in {fig:wloop} {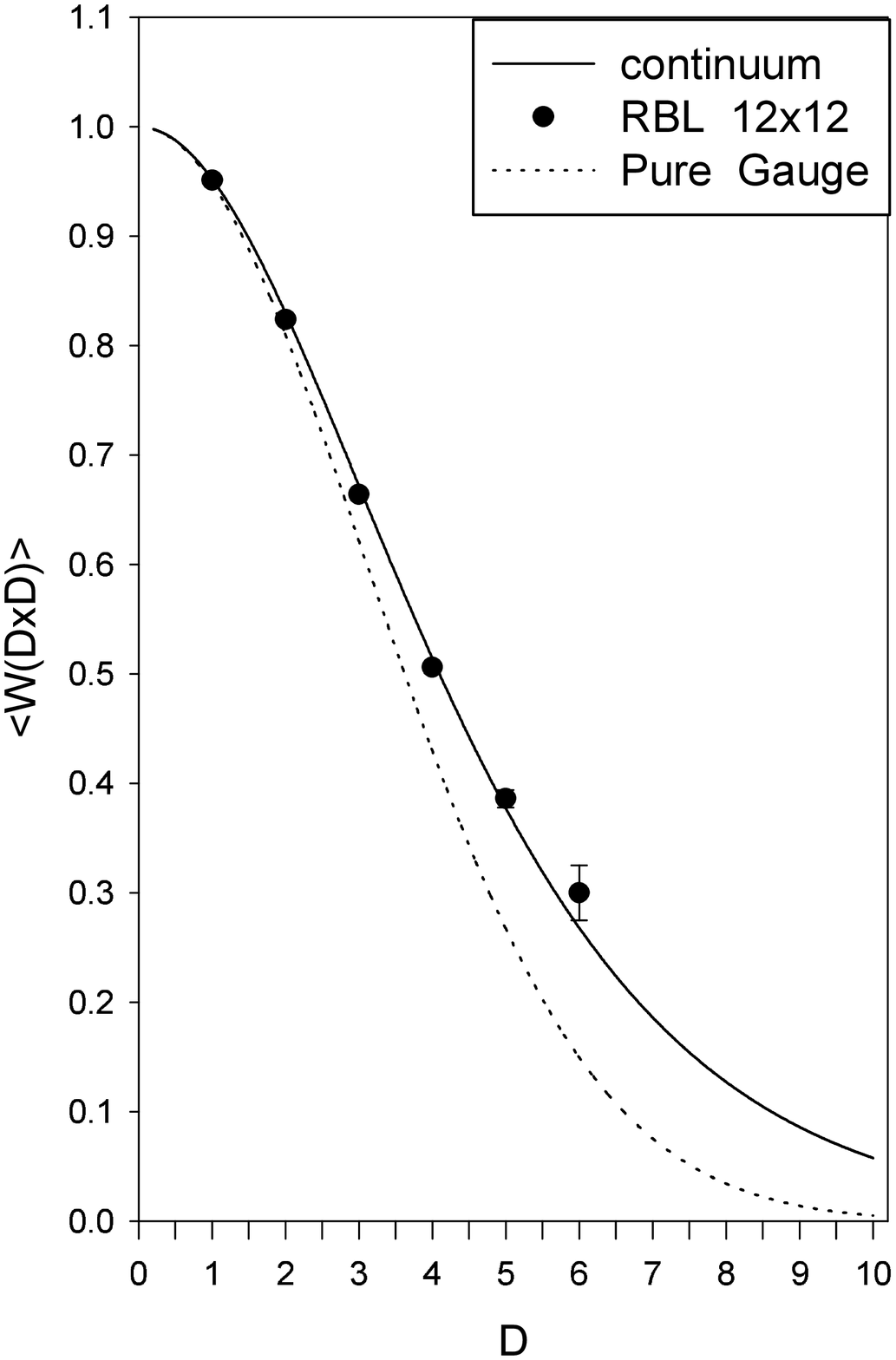} {
The values for $ D \times D $ Wilson loop plotted vs. $ D $ at
$ \beta=10.0 $. The solid line corresponds to the continuum Schwinger model
while the dotted line to the pure gauge theory.
}

\section{CONCLUSIONS and DISCUSSIONS}

The present investigation of the dynamical fermion field on RBL is far
from completed yet. Besides the Wilson loop and the chiral correlation
function, other observables must also be measured, all must be with higher
statistics and on larger lattices as well as for many different $ \beta $
values. These measurements are now in progress and the results will be
presented elsewhere. The present preliminary results are consistent
with the conclusions drawn from previous studies of free fermion and
fermion in a background gauge field on RBL, that an ensemble of RBL could
provide a proper measure for functional integrals involving fermion fields.
Without breaking the chiral symmetry at the tree level, RBL regularization
uses the naive fermion which presumably suffers from the species doubling
on any one of the lattices, however miraculously gives the correct result
when the functional integrals are sumed over all RBL. The most remarkable
feature is that it produces the correct chiral symmetry breaking through
quantum corrections. Whether RBL regularization also works for the chiral
guage theories, say chiral Schwinger model or the Electroweak theory
is still an open but very interesting question.

\end{document}